\documentclass[prl,twocolumn,showpacs,superscriptaddress]{revtex4}

\usepackage{amsmath}
\usepackage{amsfonts}
\usepackage{amssymb}
\usepackage{graphicx}
\usepackage{bm}

\begin{document}

\title{Efficient Hidden-Variable Simulation of Measurements in Quantum Experiments}

\author{Borivoje Daki\' c}
\affiliation{Institute for Quantum Optics and Quantum Information,
Austrian Academy of Sciences, Boltzmanngasse 3, A-1090 Vienna, Austria}
\affiliation{Faculty of Physics, University of Vienna, Boltzmanngasse 5, A-1090 Vienna, Austria}

\author{Milovan \v Suvakov}
\affiliation{Institute of Physics, Pregrevica 118, 11080 Belgrade, Serbia}

\author{Tomasz Paterek}
\affiliation{Institute for Quantum Optics and Quantum Information,
Austrian Academy of Sciences, Boltzmanngasse 3, A-1090 Vienna, Austria}

\author{{\v C}aslav Brukner}
\affiliation{Institute for Quantum Optics and Quantum Information,
Austrian Academy of Sciences, Boltzmanngasse 3, A-1090 Vienna,
Austria} \affiliation{Faculty of Physics, University of Vienna,
Boltzmanngasse 5, A-1090 Vienna, Austria}

\date{\today}

\begin{abstract}
We prove that the results of a finite set of general quantum measurements 
on an arbitrary dimensional quantum system can be simulated 
using a polynomial (in measurements) number of hidden-variable states.
In the limit of infinitely many measurements, our method gives models
with the minimal number of hidden-variable states, which scales linearly with the number of measurements.
These results can find applications in foundations of quantum theory, 
complexity studies and classical simulations of quantum systems.
\end{abstract}


\maketitle

In classical physics, the position and momentum of a particle determine the outcomes
of all possible measurements that can be performed upon it.
They define a deterministic classical state.
If the state is not fully accessible,
a general probabilistic classical state
is a mixture of the deterministic states, arising from the inaccessibility.
Since quantum mechanics gives only probabilistic predictions,
it was puzzling already to the fathers of the theory
whether it can be completed with an underlying classical-like model~\cite{EPR}.
The quantum probabilities would then arise from an inaccessibility of some \emph{hidden variables} (HV)
describing analogs of deterministic classical states,
the hidden-variable states,
which determine the results of all quantum measurements.

Since the seminal work of Kochen and Specker (KS), it has been known that HV models must be \emph{contextual}~\cite{KS}.
On the operational level, the contextual HV models cannot be distinguished from quantum mechanics.
However, one may ask how plausible these models are in terms of resources,
e.g., how many HV states (also called the ``\emph{ontic states}'' \cite{SPEKKENS_CONTEXTUALITY,RUDOLPH_QUANT-PH,SPEKKENS}) they require.
In addition to the fundamental question of the minimal HV model for a quantum system, 
this research is motivated by problems in quantum information theory.
In particular, HV models allow a fair comparison between complexities of quantum and classical algorithms \cite{AARONSON,HR_CONTEXTUALITY},
as a quantum algorithm can now be represented by a classical circuit.

For an infinite number of measurement settings, already a \emph{single} qubit requires \emph{infinitely} many HV states,
the result proved by Hardy \cite{HARDY} and, in a different context, by Montina \cite{MONTINA_PRL,MONTINA_PRA}.
However, these authors did not consider the scaling of the number of HV states with the number of measurements.
Harrigan and Rudolph found a deterministic HV model
that requires exponentially many HV states to simulate results
of the finite set of measurements on \emph{all} quantum states \cite{HR}.
Our construction also provides such models and consumes at most a \emph{polynomial} number of HV states,
bringing exponential improvement.
In the limit of infinitely many measurements, the number of HV states
for an indeterministic model scales \emph{linearly} with the number of measurements.
Moreover, the number of real parameters that specify these HV states 
saturates the lower bound derived by Montina \cite{MONTINA_PRA}
and, consequently, is the minimal number possible.
Our method also allows a universal generalization of the Spekkens model \cite{SPEKKENS}.

Consider a finite number, $N$, of projective measurements on a $d$-level quantum system in a state $\rho$.
The probability to observe the $r$th result in the $n$th measurement
is $p^{(n)}_r(\hat \rho)=\mathrm{Tr} [ \hat \rho \hat \Pi^{(n)}_r ]$,
where $\hat \Pi^{(n)}_r$ is a projector on the $r$th orthogonal state of the $n$th measurement,
i.e., $r=1,...,d$ and $n = 1,...,N$.
We form a $d$-dimensional vector,
$\mathbf{p}^{(n)} = (p^{(n)}_1,\dots,p^{(n)}_d)^T$,
composed of the probabilities for distinct outcomes in the $n$th measurement.
For the set of measurements, we build a $dN$-dimensional
\emph{preparation vector}, $\mathbf{p} = (\mathbf{p}^{(1)},...,\mathbf{p}^{(N)})^T$ \cite{HARDY_AXIOMS}.
The deterministic HV states predetermine the results of \emph{all} measurements
and can be represented as a $dN$-dimensional vector
\begin{equation}
\mathbf{O}_{r_1\dots r_N}=(0,\dots,1,\dots,0|\dots|0,\dots,1,\dots,0)^{T},
\label{DET_HV_STATE}
\end{equation}
where $r_n$ is the position of $1$ in the $n$th sequence 
($r_n=0,\dots,d-1$ indicates that outcome $r_n$ occurs in the $n$th measurement). 
The space of all HV states, $\Lambda$, is formed by
classical mixtures of $d^N$ deterministic states $\mathbf{O}_{r_1\dots r_N}$.

A set of $\kappa$ quantum states $\rho_1,\dots,\rho_{\kappa}$ has a HV model for $N$ measurements,
if one can find $L$ vectors $\mathbf{O}_1,\dots,\mathbf{O}_{L}\in\Lambda$ 
such that
\begin{equation}
\label{OF}
\mathbf{p}(\rho_k)=\sum_{l = 1}^{L} \alpha_{l}(k) \mathbf{O}_{l},
\quad \textrm{ for all } k = 1,...,\kappa
\end{equation}
where $\alpha_{l}(k) \geq 0$ and $\sum_{l} \alpha_{l}(k) = 1$.
The model is called \emph{deterministic} if all $\mathbf{O}_{l}$
are deterministic HV states; otherwise, it is called \emph{indeterministic}. 
The model is \emph{preparation-universal}, if the HV states
simulate any physical state $\rho$,
and it is \emph{measurement-universal} if they simulate any measurement.

Formally, the set $\Lambda$ is a convex polytope in $\mathbb{R}^{dN}$ having the states $\mathbf{O}_{l}$ as vertices. 
Since all probabilities satisfy $0 \leq p^{(n)}_r \leq 1$, any preparation vector $\mathbf{p}(\rho)$ lies inside this polytope and has a HV model.
We study the number of HV states required for the model.

We begin with a specific deterministic HV model for a two-level quantum system (qubit)
which we shall often refer to later on. 
An arbitrary state of a qubit can be represented as
$\hat \rho=\frac{1}{2}(\openone+ \sum_{i=1}^3 x_i \hat \sigma_i)$, where
$\hat \sigma_i$'s are the Pauli matrices and $\mathbf{x}=(x_1,x_2,x_3)^T$ is a Bloch vector, 
in a unit ball $|\mathbf{x}| \leq 1$. 
A set of $N$ projective measurements, with $2N$ outcomes
(states on which the qubit is projected), is described by
$2N$ unit vectors $\pm \mathbf{m}_1,\dots,\pm \mathbf{m}_N$ on the Bloch sphere.
The preparation vector for these directions is 
$\mathbf{p}(\mathbf{x})=(\frac{1\pm \mathbf{m}_1\mathbf{x}}{2},\dots,
\frac{1\pm \mathbf{m}_N\mathbf{x}}{2})$.
Since the probability for the measurement $- \mathbf{m}$
is fully determined by the one for the $+ \mathbf{m}$,
one can reduce (''compress'') preparation vector to
$\mathbf{p}(\mathbf{x})=(\frac{1 + \mathbf{m}_1\mathbf{x}}{2},\dots,
\frac{1 + \mathbf{m}_N\mathbf{x}}{2})$.
Similarly, the deterministic HV states are reduced to $N$-dimensional vectors 
$\mathbf{O}_{r_1\dots r_N}=(r_1,\dots,r_N)^T$, where $r_n = 0,1$. 
The (reduced) space $\Lambda$ is a hypercube in $N$ dimensions,
with $2^N$ vertices defined by these states.
By Carath\' eodory's theorem
\footnote{The Carath\' eodory's theorem
states that a point, $x$, in a convex polytope in $\mathbb{R}^n$
can be written as a convex combination of $n+1$ vertices.}
for each vector $\mathbf{p}(\mathbf{x}) = (p_1,\dots,p_N)^T$,
one can identify $N+1$ HV states the convex hull of which
contains $\mathbf{p}(\mathbf{x})$.
For a given $\mathbf{x}$, the vector $\mathbf{p}(\mathbf{x})$
can be written as a permutation of a reordered preparation vector
$\mathbf{p}^{\downarrow}(\mathbf{x})$ wherein the 
probabilities appear in increasing order, 
$p_1^{\downarrow} \leq p_2^{\downarrow} \leq \dots \leq p_N^{\downarrow}$,
and the latter can be expressed in terms of $N+1$ HV states as
\begin{equation}
\mathbf{p}^{\downarrow}(\mathbf{x}) = 
\left(
             \begin{array}{cccccc}
               0 & 1 & 0 & 0 & \cdots & 0 \\
               0 & 1 & 1 & 0 & \cdots & 0\\
               0 & 1 & 1 & 1 & \cdots & 0\\
               \vdots & \vdots & \vdots & \vdots & & \vdots\\
               0 & 1 & 1 & 1 & \cdots & 1\\
             \end{array}
          \right)
             \left(
                    \begin{array}{c}
                      \alpha_0 \\ \alpha_1 \\ \alpha_2 \\ \vdots \\ \alpha_{N-1} \\ \alpha_{N}
                    \end{array}
           \right),
\label{NFAC}
\end{equation}
where the columns of the displayed matrix are the HV states.
The expansion coefficients are
\begin{eqnarray}
\alpha_0 & = & 1 - p_N^{\downarrow}, \qquad \alpha_1 = p_1^{\downarrow}, \nonumber \\
\alpha_n & = & p_n^{\downarrow} - p_{n-1}^{\downarrow} \quad \textrm{ for } \quad n=2,...,N,
\end{eqnarray}
and, due to the ordering of probabilities, the coefficients are all positive and sum up to $1$.
One can suitably permute the rows in matrix given by (\ref{NFAC})
to bring the probabilities in order given by $\mathbf{p}(\mathbf{x})$.
Thus, $\mathbf{p}(\mathbf{x})$ can be written as a convex combination
of $N+1$ columns (HV states) of a reordered matrix.
The number of $N+1$ states can be further reduced. 
E.g., for two equal probabilities, $p_1=p_2$, the number of HV states 
is decreased because $\alpha_2=0$.
If, say, $p_2=1-p_1$,  one can exchange $\mathbf{m}_1\to -\mathbf{m}_1$, 
such that the probabilities become equal, leading to another reduction.
Importantly, different quantum states are generally modeled by \emph{different} sets of $N+1$ HV states.

As an illustrative example, consider a model for
three complementary measurements along
$\mathbf{m}_x,\mathbf{m}_y,\mathbf{m}_z$.
We show a nonuniversal model, only for the
eigenstates of these measurements: $\pm\mathbf{m}_x,\pm\mathbf{m}_y,\pm\mathbf{m}_z$.
The corresponding preparation vectors are:
$\mathbf{p}^{(+)}_x = (1,\frac{1}{2},\frac{1}{2})$,
$\mathbf{p}^{(-)}_x = (0,\frac{1}{2},\frac{1}{2})$,
$\mathbf{p}^{(+)}_y = (\frac{1}{2},1,\frac{1}{2})$,
$\mathbf{p}^{(-)}_y = (\frac{1}{2},0,\frac{1}{2})$,
and
$\mathbf{p}^{(+)}_z = (\frac{1}{2},\frac{1}{2},1)$,
$\mathbf{p}^{(-)}_z = (\frac{1}{2},\frac{1}{2},0)$.
Applying the method of (\ref{NFAC}) 
to each of these preparation vectors,
one finds that $L=4$ HV states
are sufficient for the simulation:
$\mathbf{O}_0 = (1,1,1)^T$,
$\mathbf{O}_1 = (1,0,0)^T$,
$\mathbf{O}_2 = (0,1,0)^T$,
and $\mathbf{O}_3 = (0,0,1)^T$.
These four states,
together with their decomposition of 
the preparation vectors,
\begin{eqnarray}
\mathbf{p}^{(+)}_x=\frac{1}{2}\mathbf{O}_0+\frac{1}{2}\mathbf{O}_1,~~~\mathbf{p}^{(-)}_x=\frac{1}{2}\mathbf{O}_2+\frac{1}{2}\mathbf{O}_3,\nonumber \\
\mathbf{p}^{(+)}_y=\frac{1}{2}\mathbf{O}_0+\frac{1}{2}\mathbf{O}_2,~~~\mathbf{p}^{(-)}_y=\frac{1}{2}\mathbf{O}_1+\frac{1}{2}\mathbf{O}_3,\nonumber \\
\mathbf{p}^{(+)}_z=\frac{1}{2}\mathbf{O}_0+\frac{1}{2}\mathbf{O}_3,~~~\mathbf{p}^{(-)}_z=\frac{1}{2}\mathbf{O}_1+\frac{1}{2}\mathbf{O}_2.
\end{eqnarray}
are equivalent to the toy model of Spekkens~\cite{SPEKKENS}.

We give a constructive proof that a \emph{preparation-universal} simulation of $N$ quantum measurements on a qubit
can be achieved with the number of HV states that is polynomial in $N$.
Let $\mathcal{M}$ denote a polytope formed as a convex hull of the measurement settings, 
$\mathcal{M} = \mathrm{conv}\{\pm\mathbf{m}_1,\dots,\pm\mathbf{m}_N\}$.
Its \emph{dual polytope} is a set
\footnote{In the special case of measurement settings within a plane, we consider the dual polygon lying in that plane.},
\begin{equation}
\label{DUAL}
\mathcal{D}_\mathcal{M}=\{\mathbf{y} \in \mathbb{R}^3| -1 \leq \mathbf{m}_n \mathbf{y} \leq 1, n=1\dots N\}.
\end{equation}
The polytope $\mathcal{M}$ lies inside the Bloch sphere and its dual contains the sphere.
Therefore, every Bloch vector can be written as a convex combination of the vertices, $\mathbf{y}_{l}$,
of the dual polytope, $\mathbf{x} = \sum_{l} \alpha_{l}(\mathbf{x}) \mathbf{y}_{l}$.
The components of the measurement vector can now be decomposed as
$p_n(\mathbf{x}) = \sum_{l} \alpha_{l}(\mathbf{x}) \frac{1}{2}(1 + \mathbf{m}_n \mathbf{y}_{l})$.
According to the definition of the dual polytope, the quantity $\frac{1}{2}(1 + \mathbf{m}_n \mathbf{y}_{l}) \in [0,1]$
and can be interpreted as the $n$th component (probability) of the $l$th HV state.
Since the Bloch vectors corresponding to projections onto orthogonal states
sum up to the zero vector, the corresponding probabilities assigned by a HV state sum up to $1$, as it should be.
Thus, the set of HV states corresponding to vertices of the dual polytope
is sufficient for a preparation-universal HV model.
Note that this model can in general be indeterministic.
In such a case, each indeterministic HV state can be further reduced into at most $N-2$ deterministic HV states,
according to (\ref{NFAC}).
The reason for $N-2$, and not $N+1$, states
stems from the observation that a vertex of the dual polytope
saturates at least three of the inequalities defining the polytope
(at least three facets have to meet at each vertex),
i.e.,\ the corresponding probability is $1$ or $0$,
and reduces the number of required deterministic HV states.
Finally, the total number of HV states required for an indeterministic model is $L \leq F$,
and for a deterministic model is $L \leq (N-2)F$,
where $F$ is the number of vertices of the dual polytope
or, equivalently, the number of facets of the measurement polytope.
A convex polytope with $2N$ vertices (in three-dimensional space)
can have $N+2 \le F \le 4(N-1)$ facets \cite{MCMULLEN},
which implies that indeterministic HV models require at most
a number of HV states that is \emph{linear} in $N$,
and deterministic ones require \emph{quadratic} number of HV states.

\begin{figure}[!b]
\begin{center}
\includegraphics[width=8.8cm]{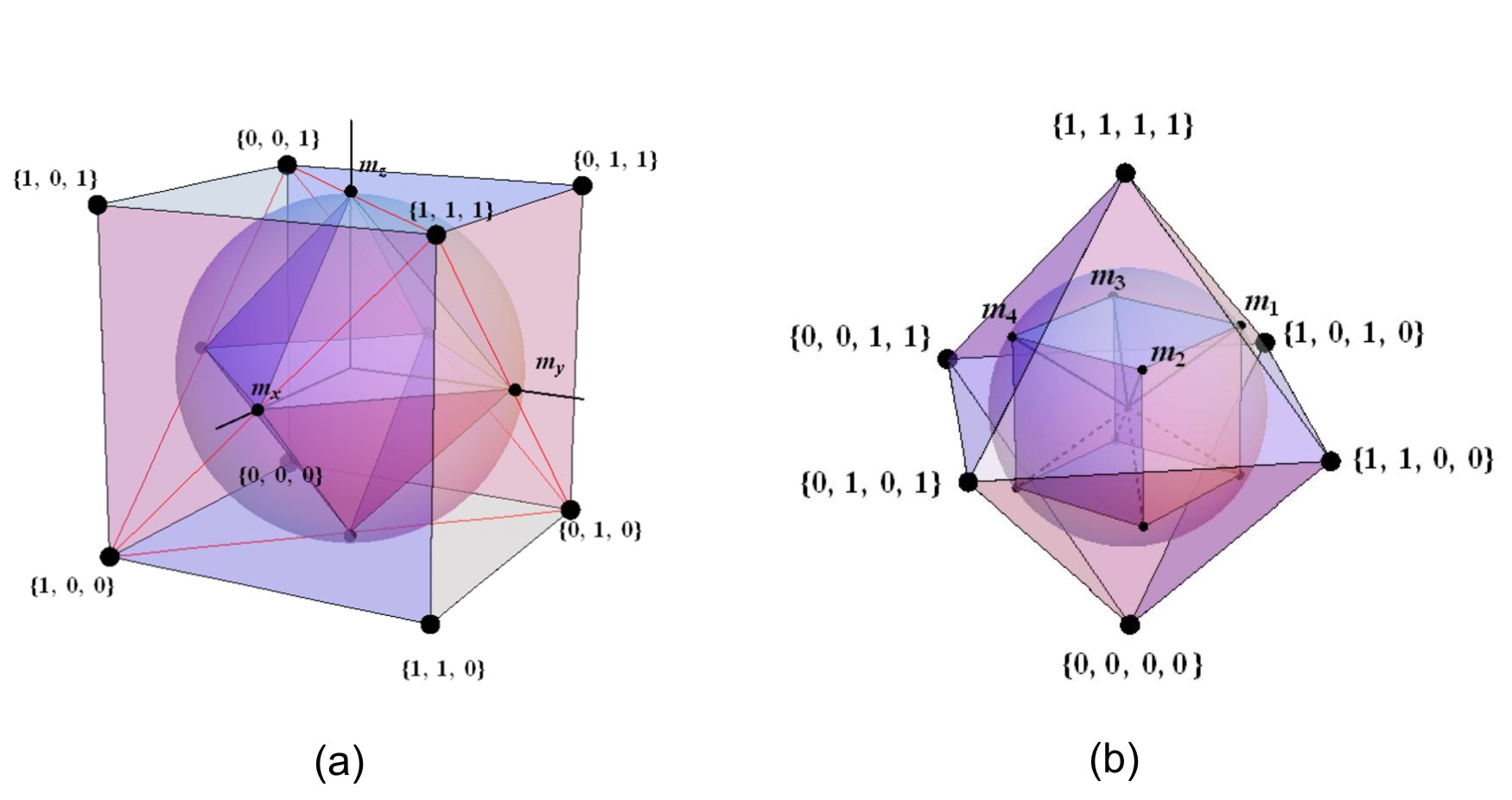}
\end{center}
\vspace{-0.5 cm}
\caption{Preparation-universal HV models and dual polytopes.
{\bf (a)} The vertices of the octahedron inside the Bloch sphere define the three complementary qubit measurements.
A \emph{preparation-universal} HV model for these measurements requires eight HV states,
which are written near their representative vertices of the cube containing the sphere.
It generalizes the Spekkens model \cite{SPEKKENS}, 
which is not universal and utilizes only four out of eight states (see main text).
Their corresponding vertices span a tetrahedron inside the cube,
which does not contain the whole Bloch sphere.
{\bf (b)} Here, the measurement directions form a cube inside the sphere.
Although more measurements are to be simulated,
the universal HV model requires only six HV states,
which are written near their representative vertices of the octahedron containing the sphere.} 
\label{FIG_OCTA}
\end{figure}

Using the dual polytope approach, we generalize Spekkens' model \cite{SPEKKENS}, 
originally formulated to explain the measurement results on the eigenstates of the three complementary directions, to the preparation-universal model.
For these directions, the measurement polytope is an octahedron, see Fig.\ \ref{FIG_OCTA}(a).
The dual polytope is a cube, whose interior forms the whole space of HV states,
with the vertices being the deterministic states.
Another interesting example is illustrated in Fig.\ \ref{FIG_OCTA}(b).

The dual polytope approach can be applied to arbitrary preparation vectors.
However, efficient simulations are only expected for highly symmetric polytopes.
For this reason, we move to more complicated Platonic solids
and general symmetry considerations.

Consider a set of measurement directions
$\pm \mathbf{m}_1$, $\dots$, $\pm \mathbf{m}_N$,
which is generated by a group;
e.g.,\ an octahedron and a cube can be generated via the chiral octahedral group $\mathcal{O}$ with $24$ rotations.
Generally, if $G$ is a symmetry of the measurement polytope, $\mathcal{M}$, 
it is also a symmetry of its dual, $\mathcal{D}_{\mathcal{M}}$;
i.e.,\ the dual polytope can also be generated by $G$.
The group action permutes the vectors $\pm \mathbf{m}_n$ as well as vertices of the dual polytope.
Since the last are related to the HV states,
we can define the permutation representation of the group in the HV space, $D_{\mathrm{P}}(G)$.
The HV state, $\mathbf{h}(\mathbf{y'})$, corresponding to a vertex of a dual polytope, $\mathbf{y}' = g \mathbf{y}$, which is generated by $g \in G$
acting on an initial vertex, $\mathbf{y}$, can be found using the group representation:
\begin{equation}
\mathbf{h}(g \mathbf{y}) = D_{\mathrm{P}}(g) \mathbf{h}(\mathbf{y}).
\label{REPR}
\end{equation}
Decomposing $\mathbf{h}(\mathbf{y})$ into deterministic HV states brings (\ref{REPR}) to the form
$\mathbf{h}(g \mathbf{y}) = \sum_{l=1} \alpha_l D_{\mathrm{P}}(g) \mathbf{O}_l$.
Therefore, the set of deterministic HV states required for the preparation-universal model is the union of a number of group orbits
$\{D_{\mathrm{P}}(g)\mathbf{O}_l | g\in G\}$.
Because of the symmetries involved, the minimal number of HV states
cannot be smaller
than the number of elements in the smallest orbit.

Let us consider two other Platonic solids, the icosahedron and the dodecahedron \footnote{Similar analysis applies to cube and octahedron.}.
Both of them posses the same symmetry, 
the chiral icosahedral group $\mathcal{I}$, with 60 rotations.
Consider the icosahedron as the measurement polytope, $N=6$.
Its dual, the dodecahedron, has $20$ vertices
corresponding to \emph{indeterministic} HV states
that can be further reduced to deterministic HV states.
The total number of possible deterministic HV states is $2^6=64$ in this case.
We have found four different orbits of action of $\mathcal{I}$ with $12,12,20,20$ different elements, respectively.
Only one orbit, with $20$ elements, gives deterministic states for universal simulation.
For $N=10$ measurement settings, the dodecahedron is the measurement polytope.
Its dual, the icosahedron, has $12$ vertices. 
The total number of possible deterministic HV states is $2^{10}=1024$,
which is partitioned into $24$ different orbits: $2$ with $12$ elements, $8$ with $20$, and $14$ with $60$ elements. 
The two lowest orbits are suitable for the universal model. 
Thus, the minimal deterministic model,
among all HV models obtained through the dual polytope construction,
requires only $24$ HV states,
twice the number of vertices of the dual polytope.

The presentation so far was limited to qubits.
However, a similar line of reasoning applies to any $d$-level quantum system.
In the general case, Pauli operators have to be replaced by generalized Gell-Mann operators, $\hat \lambda_i$,
which naturally leads to the generalized, $D \equiv d^2-1$ dimensional, Bloch representation.
An arbitrary quantum state, $\hat \rho = \frac{1}{d}[\openone + (d-1) \sum_{i=1}^{D} x_i \hat \lambda_i]$,
is now represented by a generalized Bloch vector, $\mathbf{x}$,
with components $x_i = \mathrm{Tr}(\hat \rho \hat \lambda_i)$.
We normalize the Gell-Mann operators
as $\mathrm{Tr}(\hat \lambda_i \hat \lambda_j) = \frac{d}{d-1} \delta_{ij}$,
such that pure quantum states are
represented by normalized generalized Bloch vectors.
Contrary to the qubit case, not every unit vector corresponds to a physical state.
The probability of an outcome associated with a projector on a state represented by $\mathbf{m}_n$,
in a measurement on a state represented by 
$\mathbf{x}$, is $p_n (\mathbf{x}) = \frac{1}{d}[1 + (d-1) (\mathbf{m}_n \mathbf{x})]$.
The requirement of positive probabilities reveals that, e.g., the vector $\mathbf{x} = - \mathbf{m}_n$
does not represent a physical state.

In analogy to the dual polytope,
for a set of $dN$ preparaion vectors,
representing $N$ $d$-valued observables,
we introduce a convex polytope
the interior of which includes all vectors $\mathbf{y}$
leading to physically allowed probabilities $p_n (\mathbf{y}) \in [0,1]$:
\begin{equation}
\mathcal{P}_\mathcal{M} = \{\mathbf{y} \in \mathbb{R}^D| -\tfrac{1}{d-1} \leq \mathbf{m}_n \mathbf{y} \leq 1, n=1,..., d N\}.
\label{POL}
\end{equation}
Among others, this polytope contains all the vectors of quantum states.
The generalized Bloch vectors corresponding to a complete set of orthogonal quantum states sum up to the zero vector, 
implying the probabilities assigned by a HV state for different outcomes of any measurement
sum up to $1$, as it should be.
Again, the vectors of quantum states can be expressed as a convex
combination of vertices of $\mathcal{P}_\mathcal{M}$,
and their number gives the upper bound on the amount of HV states
sufficient for preparation-universal simulation.
The polytope $\mathcal{P}_\mathcal{M}$ is specified by $q = 2dN$
linear inequalities, two inequalities for each vector $\mathbf{m}_n$,
and its maximal number of vertices
is given by $L \le {q - \delta \choose q - D} + {q - \delta' \choose q - D}$,
where $\delta \equiv \lfloor (D+1)/2 \rfloor$, $\delta' \equiv \lfloor (D+2)/2 \rfloor$,
and $\lfloor x \rfloor$ is the integer part of $x$ \cite{MCMULLEN}.
In the special case of a qubit, the dual polytope is defined by $2N$,
and not $4N$, inequalities because the two bounds of Eq.\ (\ref{DUAL}) are the same for the vectors $\pm \mathbf{m}_n$.
Since the binomial coefficient ${a \choose b}$ increases with $a$,
$L \le 2 {q - \delta \choose q - D}$.
Using ${a \choose b} = {a \choose a-b}$, we have $L \le 2 {q - \delta \choose D - \delta}$,
and since ${a \choose b} \le a^b/b!$, the maximal number of vertices is \emph{polynomial} in $N$,
$L \sim (2dN-\delta)^{D-\delta}$.
The related HV states can in general be indeterministic,
and each of them can be decomposed to $O(N)$ deterministic HV states,
using decomposition (\ref{NFAC}) in the $dN$ dimensional space $\Lambda$.
Therefore, for any system, the number of (in)deterministic HV states
required for a preparation-universal simulation
is polynomial in $N$.

In the limit of infinitely many measurements, 
our method gives (preparation and measurement) universal models with the minimal number of HV states.
As proved by Montina, in this limit the optimal model
requires $2(d-1)$ real parameters to describe the HV states \cite{MONTINA_PRA}.
We show that for an infinite number of settings
the set of universal HV states converges to the set of pure quantum states,
which is known to be parameterized by $2(d-1)$ real numbers.
First, consider a finite set of projectors $\hat \Pi_n$ with $n = 1,...,dN$,
and the corresponding polytope (\ref{POL}) in the Hilbert-Schmidt space of Hermitian operators with unit trace.
The operators of its vertices, $\hat y_l$, correspond to the HV states,
i.e.,\ for all $n$, $\mathrm{Tr}(\hat y_l \hat \Pi_n)$ gives the probability
that is assigned by the HV state,
of the outcome associated with projector $\hat \Pi_n$.
For other projectors, not within the set of $dN$, the trace does not have to represent a probability
and therefore the set of operators $\hat y_l$ is larger than the set of quantum states
\footnote{E.g., if the preparation vector of a qubit involves projectors on $| z \pm \rangle$
and $| x \pm \rangle$, it is valid to consider 
$\hat y_l = \frac{1}{2}\openone + \hat \sigma_x + \hat \sigma_z$,
which is not a quantum state.}.
However, in the limit of infinitely many measurements,
$\mathrm{Tr}(\hat y_l \hat \Pi_n) \in [0,1]$ for all possible projectors;
therefore, the eigenvalues of $\hat y_l$'s lie within the $[0,1]$ interval.
Since $\mathrm{Tr}(\hat y_l) = 1$, the operators $\hat y_l$ are just quantum states
and the HV states corresponding to pure quantum states are universal.
Their number scales \emph{linearly} with $N$, because 
$N$ measurements correspond to $dN$ projectors and each of them
represents one HV state (and also one pure quantum state).

Regarding the polytope $\mathcal{P}_\mathcal{M}$ in the space of Hermitian operators
allows for an easy generalization of our approach to POVM measurements.
POVM elements, $\hat E_n$, are positive operators
being vertices of a measurement polytope.
The polytope $\mathcal{P}_\mathcal{M}$ includes all the unit-trace operators
$\hat y$ for which $\mathrm{Tr}(\hat y \hat E_n) \in [0,1]$.
Since for all quantum states $\mathrm{Tr}(\hat \rho \hat E_n) \in [0,1]$,
the polytope $\mathcal{P}_\mathcal{M}$ contains all of them
and, as before, its vertices define HV states.

For a $d$-level system the KS argument
disqualifies non-contextual HV theories \cite{KS},
and one might wonder how contextuality enters our models.
Consider the KS argument of Peres \cite{PERES}.
It involves $33$ different vectors in $\mathbb{R}^3$,
which belong to $16$ different orthogonal triads.
Non-contextuality requires a value associated with a single vector to be the same
irrespectively of other vectors in the triad.
In the present models, the results of $16$ different measurements
are described by HV states with $3 \cdot 16 = 48$ components;
i.e.,\ a value assigned to the same vector can depend on 
the other vectors in the triad.

In conclusion, we proved that a preparaion-universal HV model of 
the results of $N$ quantum measurements requires at most 
a number of HV states which is polynomial in $N$.
In the limit of infinitely many measurements, our method gives optimal 
preparation- and measurement-universal HV models,
with the minimal number of real parameters describing the HV states.
There is no HV model that would require less HV states
than the model in which every quantum state is associated with a HV state \cite{MONTINA_PRA}.
Furthermore, since there are infinitely many measurements that can be performed on a quantum system,
its HV description requires infinitely many HV states.
This ``ontological baggage'' \cite{HARDY} 
can be seen as an argument against the HV approach
because it is extremely resource demanding already for a single qubit.

\emph{Acknowledgments}.
This work is supported by the FWF Project No. P19570-N16,
EC Project QAP (No. 015846),
the FWF project CoQuS (No. W1210-N16),
and by the Foundational Questions Institute (FQXi).

\end{document}